\documentclass[aps, pre, reprint]{revtex4-1} 
\usepackage{graphicx}
\usepackage{dcolumn}
\usepackage{amsmath}
\usepackage{amssymb}
\usepackage{color}
\usepackage{bm}

\begin{document}
\title{Divergence of the Floquet--Magnus expansion in a periodically driven \\ one-body system with energy localization}
\author{Taiki Haga}
\email[]{haga@cat.phys.s.u-tokyo.ac.jp}
\affiliation{Department of Physics, University of Tokyo, 7-3-1, Hongo, Bunkyo-ku, Tokyo, 113-8654, Japan}

\begin{abstract}
The Floquet--Magnus expansion is a useful tool to calculate an effective Hamiltonian for periodically driven systems.
In this study, we investigate the convergence of the expansion for a one-body nonlinear system in a continuous space, a driven anharmonic oscillator.
In this model, all eigenstates of the time evolution operator are found to be localized in energy space, and the expectation value of the energy is bounded from above.
We first propose a general procedure to estimate the radius of convergence of the Floquet--Magnus expansion for periodically driven systems with an unbounded energy spectrum.
By applying it to the driven anharmonic oscillator, we numerically show that the expansion diverges for all driving frequencies even if the anharmonicity is arbitrarily small.
This conclusion contradicts the widely accepted belief that the divergence of the Floquet--Magnus expansion is a direct consequence of quantum ergodicity, which implies that each eigenstate of the time evolution operator is a linear combination of all available eigenstates of the unperturbed Hamiltonian and the system heats up to infinite temperature after long intervals.
\end{abstract}

\maketitle

\section{Introduction}
\label{sec:Introduction}

Periodically driven systems have recently attracted considerable interest as a useful platform to investigate novel types of quantum many-body states \cite{Goldman-14,Bukov-15,Eckardt-17}.
In particular, high-frequency driving allows us to realize spin-orbit coupling \cite{Anderson-13} and artificial gauge fields in neutral atoms \cite{Miyake-13,Aidelsburger-13}, and also to control the topological properties of materials \cite{Jotzu-14,Goldman-16}.
According to the Floquet theory, the stroboscopic dynamics of periodically driven systems is described by a time-independent effective Hamiltonian.
The Floquet--Magnus expansion provides a formal expansion of the effective Hamiltonian with respect to the period of the driving \cite{Blanes-09,Casas-01,Mananga-11,Mananga-16}.
It is believed that if the driving frequency is much higher than the characteristic timescale of the system, the effective Hamiltonian is approximately given by a finite truncation of the expansion.
Since it is impossible to obtain the exact effective Hamiltonian for generic periodically driven many-body systems, the Floquet--Magnus expansion has been the only analytical technique to investigate the effect of the high-frequency driving, except for the flow equation approach proposed recently \cite{Vogl-19}.

Although the application of the Floquet--Magnus expansion to periodically driven quantum systems has a long history, little is understood about its radius of convergence.
The divergence of the expansion is believed to be closely related to the resonant behavior of the system.
For example, let us consider the harmonic oscillator driven by a periodic force.
It can be shown that the radius of convergence of the Floquet--Magnus expansion is equal to the natural period of the oscillator \cite{Fernandez-88,Fernandez-90,Salzman-86}.
This means that the expansion diverges when the system indefinitely absorbs the energy from the driving owing to the resonance.
From the result of the driven harmonic oscillator, one may be led to a general conjecture which states that the Floquet--Magnus expansion converges if the driving frequency is larger than any resonant frequencies of the system.
Indeed, for periodically driven two-level systems, the radius of convergence is inversely proportional to the energy level separation \cite{Salzman-86,Fel'dman-84,Maricq-87}.

For nonintegrable many-body systems, there is no established argument to connect the energy absorption of the system with the divergence of the Floquet--Magnus expansion.
However, in several previous works it was concluded that its radius of convergence for nonintegrable many-body systems vanishes in the thermodynamic limit from the fact that these systems eventually heat up to infinite temperature for an arbitrary driving frequency \cite{D'Alessio-13,Lazarides-14,D'Alessio-14,Ponte-15-1,Moessner-17,Bukov-16}.
In these cases, each Floquet state, which is the eigenstate of the time evolution operator for a single period, is a linear combination of all available eigenstates of the unperturbed Hamiltonian.
This delocalization of the Floquet states in energy space can be explained by the resonances in many-body spectrum \cite{Bukov-16}.
In the presence of strong disorder, there exists a critical driving frequency at which a transition between a many-body localized phase and an ergodic phase takes place \cite{Ponte-15-2,Lazarides-15}.
It has been believed that this critical driving frequency is identical to the radius of convergence of the Floquet--Magnus expansion.
However, a detailed study to clarify whether the divergence of the Floquet--Magnus expansion necessarily results in the heat up of the system is still lacking because it is difficult to estimate its radius of convergence for nonintegrable many-body systems.

In this study, we consider a periodically driven one-body system whose classical counterpart exhibits chaotic behavior.
The simplicity of such a system allows us to investigate the relation between the energy absorption of the system and the divergence of the Floquet--Magnus expansion.
Especially, we restrict our focus to a periodically driven oscillator with a quartic potential.
The Floquet states of this system are exponentially localized in energy space and the expectation value of the energy remains finite despite the unboundedness of the energy spectrum.
We attempt to numerically estimate the radius of convergence of the Floquet--Magnus expansion for this driven anharmonic oscillator.
The difficulty of this problem is that the Hamiltonian is an unbounded operator in an infinite-dimensional Hilbert space.
If the dimension of the Hilbert space is finite, the calculation of the radius of convergence is straightforward.
However, for a given formal expansion of an unbounded operator, even the definition of its convergence is not trivial.
In the first part of this paper, we propose a general procedure to estimate the radius of convergence of the Floquet--Magnus expansion for periodically driven systems with an unbounded energy spectrum.
This procedure is summarized as follows.
(i) The infinite-dimensional Hilbert space is truncated up to a finite dimension by using an appropriate set of orthogonal basis vectors.
(ii) The Floquet--Magnus expansion is calculated for the truncated Hamiltonian.
(iii) For each matrix element of the expansion, the infinite limit of the cutoff dimension is taken.
(iv) The radius of convergence is then defined by the ratio between successive orders of the expansion for each matrix element.
This method can reproduce the correct result for the driven harmonic oscillator.
In the second part of this paper, by applying this procedure to the driven anharmonic oscillator, we show that the radius of convergence vanishes, while the energy of the system is bounded from above.
From this result, we conjecture that the divergence of the Floquet--Magnus expansion is a universal feature of periodically driven nonlinear systems with energy localization.
In previous works \cite{D'Alessio-13,Lazarides-14,D'Alessio-14,Ponte-15-1,Moessner-17,Bukov-16}, it was widely believed that the divergence of the expansion is a consequence of quantum ergodicity, which implies that the time evolution operator is described by a random matrix and the system heats up to infinite temperature after long intervals.
Our result requests reconsideration of this consensus.
Finally, we attempt to explain the vanishing of the radius of convergence by assuming the presence of infinitely many quantum resonances.

The remainder of this paper is organized as follows.
In Sec.~\ref{sec:Floquet_Magnus_expansion}, we explain a procedure to estimate the radius of convergence of the Floquet--Magnus expansion for periodically driven systems with an unbounded energy spectrum.
In Sec.~\ref{sec:Models}, we define several models which will be considered in the following sections, harmonic oscillators driven by an external force or a parametric modulation of the natural frequency, and a driven oscillator with a quartic potential.
In Sec.~\ref{sec:Driven_anharmonic_oscillator}, we show that each Floquet state of the driven anharmonic oscillator is a linear combination of a finite number of the eigenstates of the harmonic oscillator.
This implies that the expectation value of the energy is bounded.
We also show that the level spacing statistics of the Floquet operator follows the Poisson distribution, not the Wigner-Dyson distribution, which is a consequence that the Floquet states are localized in energy space.
In Sec.~\ref{sec:Radius_of_convergence}, we first demonstrate that the procedure proposed in Sec.~\ref{sec:Floquet_Magnus_expansion} can reproduce the correct results for the driven harmonic oscillators.
We next estimate the radius of convergence for the driven anharmonic oscillator and show that it vanishes for arbitrarily small anharmonicity.
In Sec.~\ref{sec:Quantum_resonance}, we discuss the possibility that the existence of infinitely many quantum resonances leads to the divergence of the Floquet--Magnus expansion.
In Sec.~\ref{sec:Conclusions}, we provide a summary of this study and an outlook to open problems.

\section{Floquet--Magnus expansion for an unbounded Hamiltonian}
\label{sec:Floquet_Magnus_expansion}

Let us consider a periodically driven one-body system,
\begin{equation}
\hat{H}(t) = \hat{H}_0 + \lambda(t) \hat{H}_1,
\label{H_total}
\end{equation}
\begin{eqnarray}
\hat{H}_0 = \frac{1}{2} \hat{p}^2 + V_0(\hat{x}), \:\:\: \hat{H}_1 = V_1(\hat{x}),
\label{Def_H}
\end{eqnarray}
where $\hat{p}=-i\partial/\partial x$ is the momentum operator.
Here, $\lambda(t)$ is a mean-zero periodic function with period $T$, for example, $\cos(2\pi t/T)$.
As we are interested in the behavior of the system as a function of the driving period $T$, it is convenient to write $\lambda(t)=\tilde{\lambda}(t/T)$, where $\tilde{\lambda}(\tau)$ is a fixed protocol with period one, for example, $\cos(2\pi \tau)$.
We assume that the unperturbed Hamiltonian $\hat{H}_0$ is bounded from below and its spectrum is discrete.
The time evolution operator $\hat{U}(t)$ is obtained by solving 
\begin{equation}
i \partial_t \hat{U}(t) = \hat{H}(t) \hat{U}(t),
\label{EM_U}
\end{equation}
with the initial condition $\hat{U}(0)=\hat{I}$, where $\hat{I}$ is the identity operator.
From the time evolution operator for a single period $\hat{U}(T)$, which is called the Floquet operator, the effective Hamiltonian $\hat{H}_{\mathrm{F}}$ is formally defined by
\begin{equation}
e^{-i\hat{H}_{\mathrm{F}}T} = \hat{U}(T).
\label{Def_H_F}
\end{equation}
Note that $\hat{H}_{\mathrm{F}}$ is not uniquely determined only by Eq.~(\ref{Def_H_F}).
To avoid this ambiguity, we assume that $\hat{H}_{\mathrm{F}}$ is a continuous function of the driving period $T$ and, at $T=0$, it is identical to the unperturbed Hamiltonian $\hat{H}_0$.
From the above definition, the stroboscopic dynamics of the periodically driven system is described by a time-independent effective Hamiltonian $\hat{H}_{\mathrm{F}}$, which is also called the Floquet Hamiltonian.

If the dimension of the Hilbert space is finite, the Floquet Hamiltonian always exists because an arbitrary unitary matrix $U$ can be expressed as $U=e^{iH}$ in terms of a Hermitian matrix $H$.
However, if the dimension of the Hilbert space is infinite, the existence of the Floquet Hamiltonian is not guaranteed.
For example, let us consider the harmonic oscillator driven by a periodic force,
\begin{equation}
\hat{H}(t) = \frac{1}{2} \hat{p}^2 + \frac{1}{2} \omega_0^2 \hat{x}^2 + g\hat{x} \cos \omega t.
\label{H_HO_cos}
\end{equation}
The Floquet Hamiltonian is exactly given by 
\begin{equation}
\hat{H}_{\mathrm{F}} = \frac{1}{2} \hat{p}^2 + \frac{1}{2} \omega_0^2 \hat{x}^2 + g\hat{x} \frac{\omega_0^2}{\omega_0^2-\omega^2},
\label{HF_HO}
\end{equation}
(see Ref.~\cite{Fernandez-88} for a detailed derivation).
Thus, if the driving frequency $\omega$ is equal to the natural frequency $\omega_0$, the Floquet Hamiltonian does not exist.

We discuss the definition of the Floquet Hamiltonian for periodically driven systems with an unbounded energy spectrum.
Let $\{ |\phi_i \rangle \}_{i=0,1,...}$ be a set of orthogonal basis vectors spanning the infinite-dimensional Hilbert space.
We introduce a cutoff dimension $D$ and truncate the Hilbert space by using a finite set of the basis vectors $\{ |\phi_i \rangle \}_{i=0,1,...,D-1}$.
The truncated Hamiltonian matrix is defined by
\begin{eqnarray}
H^{(D)}(t)_{ij} &=& H_{0,ij}^{(D)}+\lambda(t)H_{1,ij}^{(D)} \nonumber \\
&=& \langle \phi_i|\hat{H}_0|\phi_j \rangle + \lambda(t) \langle \phi_i|\hat{H}_1|\phi_j \rangle,
\end{eqnarray}
where the indices $i,j$ run from $0$ to $D-1$.
From Eq.~(\ref{EM_U}) with the $D$-dimensional matrix $H^{(D)}(t)$, one obtain the time-evolution operator for a single period $U^{(D)}(T)$.
Then, there is a unique Floquet Hamiltonian
\begin{equation}
H_{\mathrm{F}}^{(D)} = \frac{i}{T} \log U^{(D)}(T),
\end{equation}
satisfying
\begin{equation}
\lim_{T \to 0} H_{\mathrm{F}}^{(D)} = H_0^{(D)}.
\end{equation}
Each matrix element of $H_{\mathrm{F}}^{(D)}$ is an analytic function of the driving period $T$.
For a fixed index $(i,j)$, $H_{\mathrm{F},ij}$ is defined by
\begin{equation}
H_{\mathrm{F},ij} = \lim_{D \to \infty} H_{\mathrm{F},ij}^{(D)},
\label{H_F_lim_D}
\end{equation}
and then, the Floquet Hamiltonian is given by
\begin{equation}
\hat{H}_{\mathrm{F}} = \sum_{i,j=0}^{\infty} H_{\mathrm{F},ij} |\phi_i \rangle \langle \phi_j |.
\end{equation}
In this definition, $\hat{H}_{\mathrm{F}}$ obviously depends on the choice of the orthogonal basis vectors used to truncate the Hilbert space.
However, we expect that $\hat{H}_{\mathrm{F}}$ is independent of the basis if it is chosen ``appropriately''.
One of the natural choices is the eigenstate of the unperturbed Hamiltonian,
\begin{equation}
\hat{H}_0|\phi_i^{(0)} \rangle = E_i^{(0)}|\phi_i^{(0)} \rangle, \:\:\: (i=0,1,...),
\label{H_0_eigenstates}
\end{equation}
where $E_i^{(0)}$ is the corresponding eigenvalue.
The eigenstates are sorted according to their eigenvalues, $E_i^{(0)} \leq E_{i+1}^{(0)}$.
At the end of this section, we discuss the condition of the ``good'' basis for the truncation of the Hilbert space.
An important question is whether the limit Eq.~(\ref{H_F_lim_D}) exists.
For example, in the case of the driven harmonic oscillator given by Eq.~(\ref{H_HO_cos}), this limit should not exist when $T=2\pi/\omega_0$.
In general, if the energy of periodically driven systems indefinitely increases, the Floquet Hamiltonian $H_{\mathrm{F}}$ does not exist.

We formally expand the Floquet Hamiltonian with respect to the driving period,
\begin{equation}
\hat{H}_{\mathrm{F}} = \sum_{n=0}^{\infty} \hat{\Omega}_n T^n.
\end{equation}
The first two terms are given by
\begin{equation}
\hat{\Omega}_0 = \frac{1}{T} \int_0^T dt \hat{H}(t),
\end{equation}
\begin{equation}
\hat{\Omega}_1 = \frac{1}{2iT^2} \int_0^T dt_1 \int_0^{t_1} dt_2 [\hat{H}(t_1),\hat{H}(t_2)],
\end{equation}
where $[\hat{A},\hat{B}]=\hat{A}\hat{B}-\hat{B}\hat{A}$.
The $n$-th term $\hat{\Omega}_n$ involves the $(n+1)$-nested commutators of the Hamiltonian at different times, $[\hat{H}(t_1),[\hat{H}(t_2),...,[\hat{H}(t_{n}),\hat{H}(t_{n+1})]...]]$.
This expansion is known as the Floquet--Magnus expansion \cite{Blanes-09,Casas-01,Mananga-11,Mananga-16}.
From $\hat{\Omega}_0=\hat{H}_0$, in the limit $T \to 0$, $\hat{H}_{\mathrm{F}}$ is reduced to the unperturbed Hamiltonian $\hat{H}_0$.
Note that each term $\hat{\Omega}_n$ depends on the choice of the initial phase of the periodic protocol $\lambda(t)$ (see Ref.~\cite{Goldman-14}).

If the dimension of the Hilbert space is finite, it is rigorously shown that the Floquet--Magnus expansion converges if
\begin{equation}
\int_0^T \| \hat{H}(t) \| dt < \pi,
\end{equation}
where the matrix norm is defined by $\| \hat{A} \|=\sqrt{\mathrm{tr}\hat{A}\hat{A}^{\dag}}$ (see Ref.~\cite{Blanes-09,Moan-08}).
From this theorem, if we define the typical band width of the Hamiltonian
\begin{equation}
W = \frac{1}{T} \int_0^T [E_{\mathrm{max}}(t) - E_{\mathrm{min}}(t)] dt,
\end{equation}
where $E_{\mathrm{max}}(t)$ and $E_{\mathrm{min}}(t)$ are the maximum and minimum eigenvalues of $\hat{H}(t)$, respectively, the radius of convergence $T_{\mathrm{c}}$ is then expected to behave as
\begin{equation}
T_{\mathrm{c}} \sim W^{-1}.
\label{Tc_W}
\end{equation}
This rough estimation is consistent with the results for periodically driven two-level systems \cite{Salzman-86,Fel'dman-84,Maricq-87}.
Note that $\hat{H}_{\mathrm{F}}$ does not necessarily exhibit a singular behavior at $T=T_{\mathrm{c}}$.
In fact, if the dimension of the Hilbert space is finite, there exists an analytic $\hat{H}_{\mathrm{F}}$ in the whole $T$ region, whereas the radius of convergence $T_{\mathrm{c}}$ is always finite.
This is because $\hat{H}_{\mathrm{F}}$ has poles in the complex $T$-plane out of the real axis and $T_{\mathrm{c}}$ is equal to the minimal distance between the origin and these poles.
In general, $T_{\mathrm{c}}$ is given by the ratio between the two successive terms,
\begin{equation}
T_{\mathrm{c}} = \lim_{n \to \infty} \frac{\| \hat{\Omega}_n \| }{ \| \hat{\Omega}_{n+1} \| }.
\label{Tc_ratio}
\end{equation}

If the dimension of the Hilbert space is infinite and the Hamiltonian operator is unbounded, the definition of the radius of convergence is not trivial.
As the matrix norm $\| ... \|$ is ill-defined in such cases, Eq.~(\ref{Tc_ratio}) does not make sense.
Thus, as in the definition of $\hat{H}_{\mathrm{F}}$, we truncate the Hilbert space up to a finite dimension $D$ by using the eigenstates of the unperturbed Hamiltonian $\{ |\phi_i^{(0)} \rangle \}_{i=0,1,...}$.
The Floquet--Magnus expansion for the truncated Floquet Hamiltonian is written as
\begin{equation}
H_{\mathrm{F}}^{(D)} = \sum_{n=0}^{\infty} \Omega_n^{(D)} T^n,
\label{truncated_FM_expansion}
\end{equation}
and its radius of convergence is given by
\begin{equation}
T_{\mathrm{c}}^{(D)} = \lim_{n \to \infty} \frac{\| \Omega_n^{(D)} \| }{ \| \Omega_{n+1}^{(D)} \| }.
\label{Tc_ratio_D}
\end{equation}
Although one may expect that $T_{\mathrm{c}}=\lim_{D \to \infty} T_{\mathrm{c}}^{(D)}$, it is not correct.
Note that the typical band width of the truncated Hamiltonian increases with $D$, e.g., $W \propto D$ for the harmonic oscillator, thus from Eq.~(\ref{Tc_W}), we find that $T_{\mathrm{c}}^{(D)}$ always vanishes in the limit $D \to \infty$.
In particular, this definition of $T_{\mathrm{c}}$ cannot reproduce the analytic result $T_{\mathrm{c}}=2\pi/\omega_0$ for the driven harmonic oscillator.

Although the matrix norm of $\Omega_n$ is ill-defined for an unbounded Hamiltonian, its matrix element $(\Omega_n)_{ij}$ is well-defined for appropriate basis vectors.
In fact, for a driven anharmonic oscillator which will be defined in the next section, if one employs the eigenstates of the harmonic oscillator as the basis vectors, it is possible in principle to calculate the matrix element $(\Omega_n)_{ij}$ without the truncation of the Hamiltonian.
Then, it is natural to define $T_{\mathrm{c}}$ as the ratio between $|(\Omega_n)_{ij}|$ and $|(\Omega_{n+1})_{ij}|$.
However, as the order of the expansion $n$ increases, the exact calculation of the matrix element $(\Omega_n)_{ij}$ becomes exponentially harder.
In contrast, for a finite-dimensional Hamiltonian there is an efficient algorithm to calculate the Floquet--Magnus expansion \cite{Klarsfeld-89}.
Thus, we truncate the Hamiltonian up to the cutoff dimension $D$, and after the calculation of each matrix element $(\Omega_n^{(D)})_{ij}$, take the infinite limit of $D$,
\begin{equation}
(\Omega_n^{\infty})_{ij} = \lim_{D \to \infty} (\Omega_n^{(D)})_{ij}.
\label{Omega_infinity_limit}
\end{equation}
If this limit exists, we define $T_{\mathrm{c}}$ by
\begin{equation}
T_{\mathrm{c}} = \lim_{n \to \infty} \frac{|(\Omega_n^{\infty})_{ij} |}{| (\Omega_{n+1}^{\infty})_{ij}|}.
\label{Tc_ratio_infinity}
\end{equation}
In Sec.~\ref{sec:Radius_of_convergence}, we show that, for the driven harmonic oscillator, the right-hand side of Eq.~(\ref{Tc_ratio_infinity}) is independent of the indices $i,j$ and equal to $2\pi/\omega_0$.
It should be noted that the limits with respect to $n$ and $D$ do not commute,
\begin{equation}
\lim_{n \to \infty} \lim_{D \to \infty} \frac{|(\Omega_n^{(D)})_{ij} |}{| (\Omega_{n+1}^{(D)})_{ij}|} \neq \lim_{D \to \infty} \lim_{n \to \infty} \frac{|(\Omega_n^{(D)})_{ij} |}{| (\Omega_{n+1}^{(D)})_{ij}|}.
\end{equation}

In the above argument, we have basically employed the eigenstates of the unperturbed Hamiltonian $\{ |\phi_i^{(0)} \rangle \}_{i=0,1,...}$ to truncate the Hilbert space.
However, in some cases, it would be more convenient to use other basis vectors.
For example, in the case of the driven anharmonic oscillator, the calculation is simplified by using the eigenstates of the harmonic oscillator as the basis.
Here, we discuss the ambiguity in the choice of the orthogonal basis vectors for the truncation of the Hilbert space.
A general criterion for the ``good'' basis is that the right-hand side of Eq.~(\ref{Omega_infinity_limit}) exists.
We expect that, as long as the limit $(\Omega_n^{\infty})_{ij}$ exists, $T_{\mathrm{c}}$ is independent of the choice of the basis.
In Sec.~\ref{sec:Radius_of_convergence}, we confirm that, for several specific models, $(\Omega_n^{\infty})_{ij}$ exists if we employ the eigenstates of the unperturbed Hamiltonian $\{ |\phi_i^{(0)} \rangle \}_{i=0,1,...}$.
We also show that, in the case of the driven anharmonic oscillator, the eigenstates of the harmonic oscillator satisfy the above condition.
This is because each eigenstate of the harmonic oscillator is close to that of $\hat{H}_0$ as long as the anharmonicity is not large.

\section{Models}
\label{sec:Models}

In this section, we define several models which will be discussed in this study.
The simplest example of integrable driven system is the harmonic oscillator driven by an external force,
\begin{equation}
\hat{H}_0 = \hat{H}_{\mathrm{HO}} = \frac{1}{2} \hat{p}^2 + \frac{1}{2} \omega_0^2 \hat{x}^2, \:\:\: \hat{H}_1 = g\hat{x}.
\label{H_HO}
\end{equation}
As mentioned in the previous section, the energy of the system diverges at the resonance $T=2\pi/\omega_0$.
We also consider a parametrically driven harmonic oscillator,
\begin{equation}
\hat{H}_0 = \hat{H}_{\mathrm{HO}}, \:\:\: \hat{H}_1 = \frac{1}{2} g\hat{x}^2.
\label{H_HO_parametric}
\end{equation}
In contrast to the harmonic oscillator driven by the periodic force, each resonance region in which the energy of the system diverges has a finite width.
In the limit $g \to 0$, the first resonance occurs at $T=\pi/\omega_0$, and the width of the unstable region around it decreases proportionally with $g$ (see Ref.~\cite{Landau}).

As a simple but nontrivial example of a periodically driven systems, we define a driven oscillator with a quartic potential,
\begin{equation}
\hat{H}_0 = \hat{H}_{\mathrm{HO}} + \frac{1}{4} \beta \hat{x}^4, \:\:\: \hat{H}_1 = g\hat{x}, \:\:\: (\beta>0).
\label{H_anharmonic}
\end{equation}
The classical and quantum dynamics of the driven anharmonic oscillator have been investigated extensively \cite{Bolotin-95,Korsch-92,Korsch-93}.
In the following, we briefly review the classical dynamics of Eq.~(\ref{H_anharmonic}).
As this is a one-dimensional system, the system is integrable without the driving force.
It is known that the energy of the system remains finite for arbitrary frequency and amplitude of the driving force.
For small amplitude $g$, the dynamics is regular for any initial condition.
When $g$ exceed a certain value, there are two threshold energies $E_1$ and $E_2$ ($E_1<E_2$) such that the system exhibits chaotic dynamics for the initial condition with $E_1<E<E_2$.
If the energy of the initial condition is lower than $E_1$ or larger than $E_2$, the dynamics is regular \cite{Bolotin-95}.

In general, for one-body and one-dimensional driven systems with a polynomial potential $x^{2n}$, $(n>1)$, it is known that the energy of the system remains finite even if the dynamics is chaotic \cite{Bolotin-95}.
This fact is understood as follows.
In the absence of the driving force, because the system is integrable, it has periodic orbits.
For the polynomial potential, the frequency of the periodic orbits increases with the energy.
In the presence of the driving force, the periodic orbits with frequencies higher than that of the driving cannot absorb the energy.
Thus, the system exhibits energy bounded chaos.
In contrast, for generic driven many-body systems, chaotic behavior is accompanied by an unbounded energy increase.
For example, let us consider a nonlinear lattice system, where $N$ anharmonic oscillators are connected by harmonic springs.
In the absence of the driving, there is a threshold energy $E_{\mathrm{c}}(N,\beta)$ such that the dynamics is regular for $E<E_{\mathrm{c}}(N,\beta)$ and chaotic for $E>E_{\mathrm{c}}(N,\beta)$ \cite{Pettini-91,Berman-05}.
Here, $E_{\mathrm{c}}(N,\beta)$ decreases to zero in the thermodynamic limit $N \to \infty$ and it diverges in the integrable limit $\beta \to 0$.
If a periodic force is applied to the chaotic state, the energy of the system is expected to increases unboundedly for arbitrary driving frequency and amplitude.
On the other hand, if the initial energy is lower than $E_{\mathrm{c}}(N,\beta)$, the energy absorption from the driving is bounded for sufficiently small driving amplitude.
In summary, one should keep in mind that chaotic dynamics without energy absorption is peculiar to one-body and one-dimensional driven systems.

\section{Energy localization in the driven anharmonic oscillator}
\label{sec:Driven_anharmonic_oscillator}

\begin{figure}
 \centering
 \includegraphics[width=0.4\textwidth]{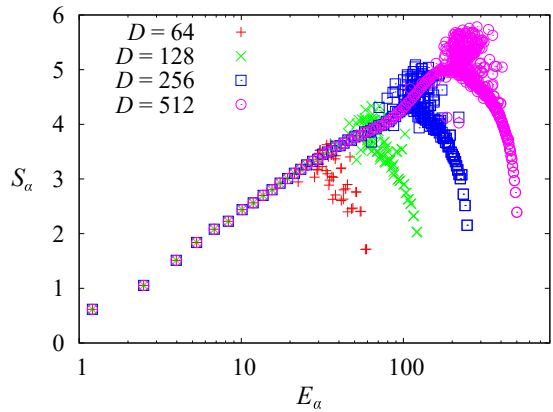}
 \caption{Plot of $S_{\alpha}$ versus $E_{\alpha} = \langle \phi_{\alpha}| \hat{H}_{\mathrm{HO}} |\phi_{\alpha} \rangle$.
 The values of the cutoff dimension are $D=64,128,256$, and $512$.
 The values of other parameters are $\omega_0=\beta=g=1$, and $T=1$.
 For each $E_{\alpha}$, the entropy $S_{\alpha}$ converges to a finite value as the cutoff dimension $D$ increases.}
 \label{fig-S}
\end{figure}

We numerically show that the Floquet state of the driven quantum anharmonic oscillator is localized in energy space and the expectation value of the energy remains finite.
As a periodic protocol $\lambda(t)$ in Eq.~(\ref{H_total}), we employ a step-like form, which has $+1$ in the first half-period and $-1$ in the second half-period,
\begin{eqnarray}
\lambda(t) = \left\{ \begin{array}{ll}
+1, & (nT \leq t < (n+1/2)T), \\
-1, & ((n+1/2)T \leq t < (n+1)T), \\
\end{array} \right.
\label{lambda_step}
\end{eqnarray}
where $n$ is an integer.
In this case, the time-evolution operator for a single period is written as
\begin{equation}
\hat{U} = e^{-i(\hat{H}_0-\hat{H}_1)T/2} e^{-i(\hat{H}_0+\hat{H}_1)T/2}.
\end{equation}
We denote the Floquet states, which are the eigenstates of $\hat{U}$, as $\{ |\phi_{\alpha} \rangle \}_{\alpha=0,1,...}$,
\begin{equation}
\hat{U} |\phi_{\alpha} \rangle = e^{-i\mu_{\alpha}} |\phi_{\alpha} \rangle,
\label{U_eigen}
\end{equation}
where $\mu_{\alpha}$ is called the quasi-energy.
By using the eigenstates of the harmonic oscillator $\{ |n \rangle \}_{n=0,1,...}$ $(\hat{H}_{\mathrm{HO}}|n \rangle=(n+1/2) \omega_0 |n \rangle)$, the Floquet state is expanded as
\begin{equation}
|\phi_{\alpha} \rangle = \sum_{n} C_{\alpha n} |n \rangle.
\end{equation}
We introduce the Shannon entropy by
\begin{equation}
S_{\alpha} = - \sum_{n} |C_{\alpha n}|^2 \ln |C_{\alpha n}|^2,
\end{equation}
which corresponds to the number of the eigenstates of the harmonic oscillator contributing to the Floquet state $|\phi_{\alpha} \rangle$.
The Hamiltonian is truncated up to a finite dimension $D$ in the eigenstates of the harmonic oscillator, and then the Floquet state is calculated by the exact diagonalization.
In Fig.~\ref{fig-S}, $S_{\alpha}$ is plotted against the expectation value of the Hamiltonian of the harmonic oscillator $E_{\alpha}= \langle \phi_{\alpha}| \hat{H}_{\mathrm{HO}} |\phi_{\alpha} \rangle$ for different values of the cutoff dimension $D$.
For a fixed $E_{\alpha}$, the entropy $S_{\alpha}$ converges to a finite value in the limit $D \to \infty$.
This implies that each Floquet state $|\phi_{\alpha} \rangle$ is a linear combination of a finite number of the eigenstates of the harmonic oscillator.

\begin{figure}
 \centering
 \includegraphics[width=0.4\textwidth]{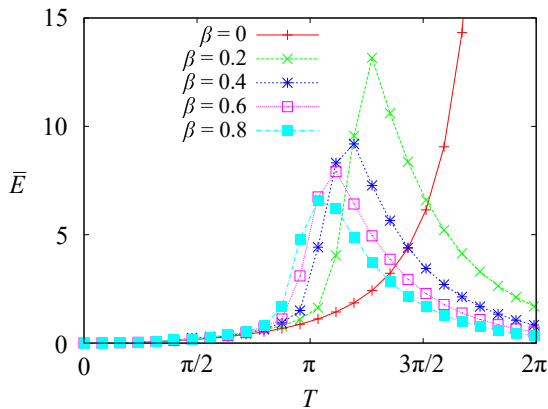}
 \caption{Long-time average of the energy $\bar{E}$ as a function of $T$.
 The values of the anharmonicity are $\beta=0$, $0.2$, $0.4$, $0.6$, and $0.8$.
 The values of other parameters are $\omega_0=g=1$.
 The cutoff dimension is $D=64$.}
 \label{fig-E}
\end{figure}

We also consider the energy absorption of the driven anharmonic oscillator directly.
A given initial state $|\psi_{0} \rangle$ is expanded as
\begin{equation}
|\psi_{0} \rangle = \sum_{\alpha} c_{\alpha} |\phi_{\alpha} \rangle.
\label{psi_0_expansion}
\end{equation}
The state at $t=nT$ is given by
\begin{equation}
|\psi_{nT} \rangle = \hat{U}^n |\psi_{0} \rangle = \sum_{\alpha} c_{\alpha} e^{-in\mu_{\alpha}} |\phi_{\alpha} \rangle.
\end{equation}
Thus, the expectation value of $\hat{H}_0$ reads
\begin{eqnarray}
E(nT) &=& \langle \psi_{nT}| \hat{H}_0 |\psi_{nT} \rangle \nonumber \\
&=& \sum_{\alpha,\alpha'} c_{\alpha}^* c_{\alpha'} e^{in(\mu_{\alpha}-\mu_{\alpha'})} \langle \phi_{\alpha}| \hat{H}_0 |\phi_{\alpha'} \rangle.
\end{eqnarray}
If there is no degeneration in $\mu_{\alpha}$, the long-time average is calculated as
\begin{eqnarray}
\bar{E} = \lim_{N \to \infty} \frac{1}{N} \sum_{n=1}^{N} E(nT) = \sum_{\alpha} |c_{\alpha}|^2 \langle \phi_{\alpha}| \hat{H}_0 |\phi_{\alpha} \rangle.
\label{E_av}
\end{eqnarray}
By using the exact diagonalization, we confirm that $\bar{E}$ is finite.
We choose the ground state of the harmonic oscillator as the initial state $|\psi_{0} \rangle$.
Figure \ref{fig-E} shows $\bar{E}$ as a function of $T$ for different values of the anharmonicity $\beta$.
The cutoff dimension is chosen sufficiently large so that $\bar{E}$ is well converged.
While for the harmonic oscillator $\bar{E}$ diverges at $T=2\pi/\omega_0$ as expected, in the presence of the anharmonicity this resonance disappears.
In Fig.~\ref{fig-E}, $\bar{E}$ is plotted for discrete values of $T$ with the interval $\Delta T=0.25$.
If we assume that $\bar{E}$ does not have other resonance peaks in a smaller scale than $\Delta T$, we can conclude that $\bar{E}$ remains finite for any $T$.
However, we cannot exclude the possibility that there exist fine resonances that are not shown in Fig.~\ref{fig-E}.
We will discuss this issue in Sec.~\ref{sec:Quantum_resonance} and show that the resonance structure of the driven anharmonic oscillator is highly complicated.

\begin{figure}
 \centering
 \includegraphics[width=0.48\textwidth]{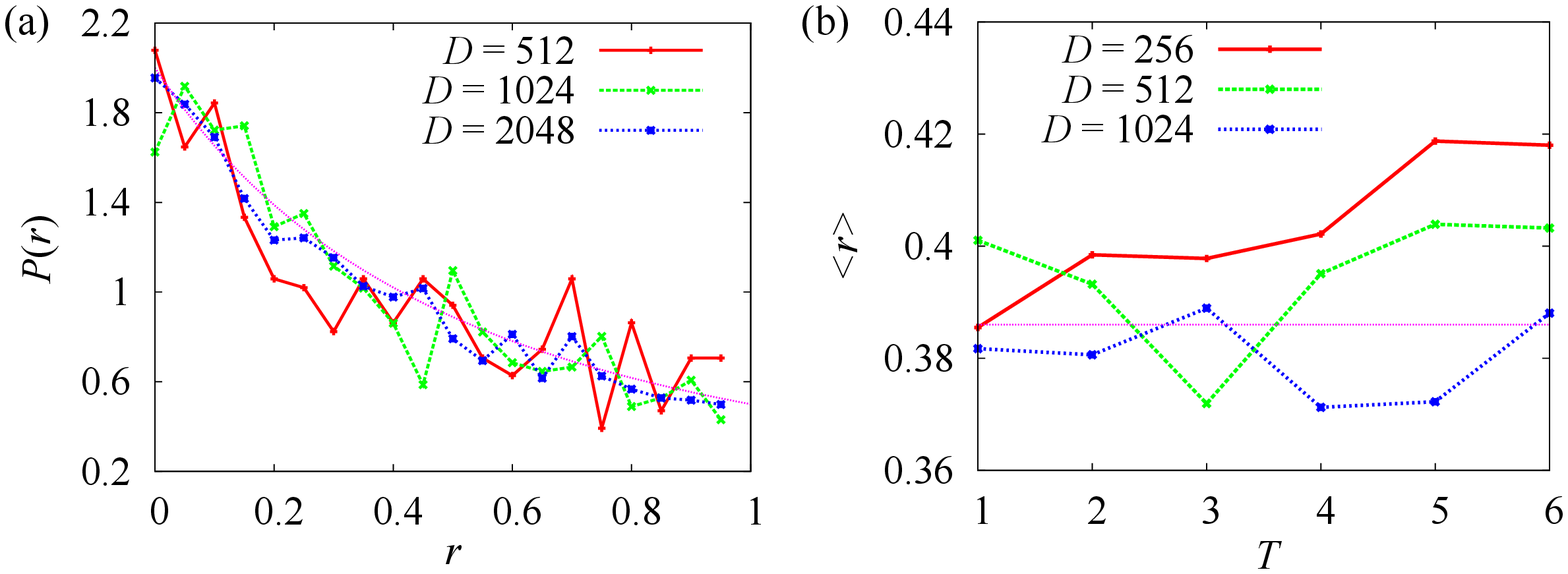}
 \caption{(a) Distribution of the normalized level spacing $r_{\alpha}$.
 The values of the cutoff dimension are $D=512,1024$, and $2048$.
 The values of other parameters are $\omega_0=\beta=g=1$, and $T=1$.
 The thin dashed line represents $P_{\mathrm{POI}}(r) = 2/(1+r)^2$.
 (b) Average of $r_{\alpha}$ as a function of $T$.
 The values of the cutoff dimension are $D=256,512$, and $1024$.
 The thin dashed line represents $\langle r\rangle_{\mathrm{POI}} \simeq 0.386$.}
 \label{fig-level}
\end{figure}

We next investigate the level spacing statistics of the Floquet operator $\hat{U}$.
It is believed that, when the Floquet operator for a periodically driven system exhibits properties of random matrices, the system heats up to infinite temperature independently of the initial states \cite{D'Alessio-14}.
This implies that, if the energy growth of the system is bounded, the level spacing of the Floquet operator is described by the Poisson distribution.
In the following, we will confirm that this is also true for the driven anharmonic oscillator.
In Eq.~(\ref{U_eigen}), we assume that the quasi-energy levels $\{ \mu_{\alpha} \}_{\alpha=0,1,...,D-1}$ are sorted in increasing order in the range $[-\pi,\pi)$.
Using $\delta_{\alpha}=\mu_{\alpha+1}-\mu_{\alpha}$, the normalized level spacing $r_{\alpha}$ ($0<r_{\alpha}<1$) is defined by
\begin{equation}
r_{\alpha} = \frac{\mathrm{min}(\delta_{\alpha},\delta_{\alpha+1})}{\mathrm{max}(\delta_{\alpha},\delta_{\alpha+1})}.
\end{equation}
If each eigenvalue of $\hat{U}$ is independently distributed on the unit circle, the level spacing $\delta_{\alpha}$ obeys the Poisson distribution.
In contrast, for driven nonintegrable many-body systems, $\delta_{\alpha}$ is shown to obey the Wigner--Dyson distribution owing to the level repulsion \cite{D'Alessio-14}.
The distributions of $r_{\alpha}$ corresponding to the Wigner--Dyson and Poisson distributions are given by
\begin{eqnarray}
P_{\mathrm{WD}}(r) &=& \frac{27}{4} \frac{r+r^2}{(1+r+r^2)^{5/2}}, \nonumber \\ P_{\mathrm{POI}}(r) &=& \frac{2}{(1+r)^2},
\label{P_theory}
\end{eqnarray}
respectively, and these averages read
\begin{equation}
\langle r \rangle_{\mathrm{WD}} \simeq 0.536, \:\:\: \langle r \rangle_{\mathrm{POI}} \simeq 0.386.
\label{r_av_theory}
\end{equation}
The left and right panels of Fig.~\ref{fig-level} show the distribution of $r_{\alpha}$ and its average as a function of $T$, respectively.
One can confirm that the level spacing distribution converges to $P_{\mathrm{POI}}(r)$ as the cutoff dimension $D$ increases.
This conclusion is consistent with the fact that the energy growth of the system is bounded.
In several previous works (see e.g. Ref.~\cite{D'Alessio-14}), it was expected that the emergence of the random matrix properties of the Floquet operator is a consequence of the divergence of the Floquet--Magnus expansion.
However, in the next section we will demonstrate that, for the driven anharmonic oscillator, the expansion diverges while the level spacing statistics follows the Poisson distribution.

Note that the driven classical anharmonic oscillator exhibits chaotic behavior depending on the parameters and initial conditions.
In general,  it is expected that the level spacing statistics of driven few-body systems with energy localization follows the Poisson distribution, even if its classical counterparts can exhibit chaotic behavior.
This fact can be understood as follows.
In such systems, each Floquet state $|\phi_{\alpha}\rangle$ is a linear combination of a small number of the eigenstates of the unperturbed Hamiltonian $\hat{H}_0$.
This implies that the overlap between almost all pairs of the Floquet states is quite small.
Thus, each quasi-energy is independently distributed in $[-\pi,\pi)$ and the level spacing follows the Poisson distribution.
This scenario is confirmed in the kicked rotor model \cite{Izrailev-90}.
The classical kicked rotor model exhibits chaotic dynamics with unbounded energy diffusion, whereas in the quantum case, the energy growth of the system eventually saturates owing to the localization of the Floquet states in the momentum space (see Sec.~\ref{sec:Quantum_resonance}).
The level spacing statistics of the kicked rotor model is known to follow the Poisson distribution.
The transition from the Poisson to the Wigner--Dyson statistics is observed in the coupled kicked rotor model, where $N$ kicked rotors are fully connected by the kicking potential.
For $N \geq 3$, this model exhibits a transition from an energy bounded state to a heating state with unbounded energy growth, as the strength of the kick exceeds a certain threshold.
It is shown that the level spacing statistics for the heating state is described by the Wigner--Dyson distribution \cite{Russomanno-18}.

\section{Radius of convergence for the driven anharmonic oscillator}
\label{sec:Radius_of_convergence}

In Sec.~\ref{sec:Floquet_Magnus_expansion}, we have proposed a procedure to determine the radius of convergence of the Floquet--Magnus expansion for periodically driven systems with an unbounded energy spectrum.
In the first part of the present section, we numerically demonstrate that this procedure can reproduce the correct results for exactly solvable models.
Let us consider the driven harmonic oscillator defined by Eq.~(\ref{H_HO}).
If we employ the step-like protocol given by Eq.~(\ref{lambda_step}), the corresponding Floquet Hamiltonian is calculated as
\begin{equation}
\hat{H}_F = \frac{1}{2} \hat{p}^2 + \frac{1}{2} \omega_0^2 \hat{x}^2 - \hat{p} \frac{g}{\omega_0} \tan \frac{\omega_0 T}{4}.
\label{HF_HO_analytic}
\end{equation}
A detailed derivation of Eq.~(\ref{HF_HO_analytic}) is given in Appendix \ref{Appendix:driven_harmonic_oscillator}.
As the radius of convergence $T_{\mathrm{c}}$ is equal to the minimal distance between the origin and poles, we have $T_{\mathrm{c}}=2\pi/\omega_0$, which is the same as the case of the monochromatic driving $\lambda(t)=\cos(2\pi t/T)$.

\begin{figure}
 \centering
 \includegraphics[width=0.48\textwidth]{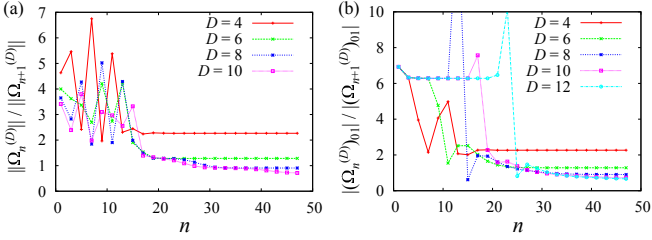}
 \caption{(a) Plot of $\| \Omega_n^{(D)} \|/\| \Omega_{n+1}^{(D)} \|$ as a function of $n$ for the driven harmonic oscillator.
The values of the parameters are $\omega_0=g=1$.
$T_{\mathrm{c}}^{(D)}=\lim_{n \to \infty} \| \Omega_n^{(D)} \|/\| \Omega_{n+1}^{(D)} \|$ decreases to zero as $D$ increases.
(b) Plot of $|(\Omega_n^{(D)})_{01} |/|(\Omega_{n+1}^{(D)})_{01} |$ as a function of $n$.
The ratio shows a plateau at small $n$.
The width of the plateau increases with $D$ and its height is equal to $2\pi$.}
 \label{fig-Tc-HO}
\end{figure}

\begin{figure}
 \centering
 \includegraphics[width=0.48\textwidth]{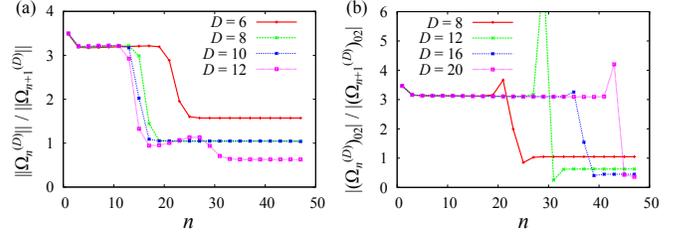}
 \caption{(a) Plot of $\| \Omega_n^{(D)} \|/\| \Omega_{n+1}^{(D)} \|$ as a function of $n$ for the parametrically driven harmonic oscillator.
The values of the parameter are $\omega_0=1$ and $g=0.1$.
(b) Plot of $|(\Omega_n^{(D)})_{02} |/|(\Omega_{n+1}^{(D)})_{02} |$ as a function of $n$.
The width of the plateau increases with $D$ and its height is nearly equal to $\pi$.}
 \label{fig-Tc-PHO}
\end{figure}

According to the procedure proposed in Sec.~\ref{sec:Floquet_Magnus_expansion}, we represent $\hat{H}(t)$ in matrix form with respect to the eigenstates of $\hat{H}_0$,
\begin{equation}
H(t)_{ij} = j\omega_0 \delta_{ij} + \frac{\lambda(t)g}{\sqrt{2\omega_0}} \bigl( \sqrt{j+1}\delta_{i,j+1}+\sqrt{j}\delta_{i,j-1} \bigr),
\end{equation}
where we have omitted the zero point energy.
We next introduce the cutoff dimension $D$ and calculate the Floquet--Magnus expansion for the truncated Hamiltonian.
The numerical scheme to calculate each term of the expansion is explained in Appendix \ref{Appendix:Floquet_Magnus_expansion}.
To confirm that the radius of convergence $T_{\mathrm{c}}^{(D)}$ defined by Eq.~(\ref{Tc_ratio_D}) does not reproduce the anticipated value $2\pi/\omega_0$, we plot $\| \Omega_n^{(D)} \|/\| \Omega_{n+1}^{(D)} \|$ versus $n$ in Fig.~\ref{fig-Tc-HO} (a).
One can see that $T_{\mathrm{c}}^{(D)}$ decreases to zero as $D$ goes to infinity.

We calculate $T_{\mathrm{c}}$ defined by Eq.~(\ref{Tc_ratio_infinity}).
For the index of the matrix elements $(i,j)$, we choose $i=0,j=1$.
In Fig.~\ref{fig-Tc-HO} (b), $|(\Omega_n^{(D)})_{01} |/|(\Omega_{n+1}^{(D)})_{01} |$ is plotted as a function of $n$ for different values of the cutoff dimension $D$.
It shows a plateau at small $n$, the width of which increases with $D$.
This implies that, in the limit $D \to \infty$, $|(\Omega_n^{\infty})_{01} |/|(\Omega_{n+1}^{\infty})_{01} |$ becomes a constant independent of $n$.
The height of the plateau is equal to $6.28 \simeq 2\pi/\omega_0$.
Thus, we have $T_{\mathrm{c}} = \lim_{n \to \infty} \lim_{D \to \infty} |(\Omega_n^{(D)})_{01} |/|(\Omega_{n+1}^{(D)})_{01} | = 2\pi/\omega_0$, which is identical to the analytic result.
We have confirmed that $T_{\mathrm{c}}$ is the same for the other indices $i=1$, $j=2$ and $i=2$, $j=3$.
At first glance, it seems strange that Fig.~\ref{fig-Tc-HO} (a) does not show any plateau at small $n$ despite the presence of the plateau in Fig.~\ref{fig-Tc-HO} (b) for an arbitrary pair $(i,j)$.
We emphasize that Eqs.~(\ref{Omega_infinity_limit}) and (\ref{Tc_ratio_infinity}) are evaluated for a fixed $(i,j)$ independent of $D$.
If we chose the pair $(i,j)$ such that it is proportional to $D$, for example, $i=j=D/2$, Eq.~(\ref{Tc_ratio_infinity}) would vanish.
Since the matrix norm $\| \Omega_n^{(D)} \|= (\sum_{i,j=0}^{D-1}|(\Omega_n^{(D)})_{ij}|^2)^{1/2}$ is dominated by the matrix element with $i \sim j \sim D$ for sufficiently large $D$, $\| \Omega_n^{(D)} \|/\| \Omega_{n+1}^{(D)} \|$ does not show any plateau at small $n$.

We next consider a parametrically driven harmonic oscillator defined by Eq.~(\ref{H_HO_parametric}).
As in the case of the driven harmonic oscillator, we calculate the Floquet--Magnus expansion for the truncated Hamiltonian.
The left and right panels of Fig.~\ref{fig-Tc-PHO} show $\| \Omega_n^{(D)} \|/\| \Omega_{n+1}^{(D)} \|$ and $|(\Omega_n^{(D)})_{02} |/|(\Omega_{n+1}^{(D)})_{02} |$ as functions of $n$, respectively.
From Fig.~\ref{fig-Tc-PHO} (a), one can confirm that $T_{\mathrm{c}}^{(D)}$ decreases to zero as $D$ goes to infinity.
In contrast, $|(\Omega_n^{(D)})_{02} |/|(\Omega_{n+1}^{(D)})_{02} |$ shows a plateau at small $n$, the width of which increases with $D$.
The height of the plateau, which is equal to $T_{\mathrm{c}}$, is slightly smaller than $\pi/\omega_0$.
Here, $T_{\mathrm{c}}$ approaches $\pi/\omega_0$ in the limit $g \to 0$.
From Fig.~\ref{fig-Tc-PHO} (a), one can see that the plateau at small $n$ becomes narrower as $D$ increases.
Although this suggests that $\lim_{D \to \infty} \| \Omega_n^{(D)} \|/\| \Omega_{n+1}^{(D)} \|=0$ for each $n$, it is not clear whether this behavior is a general feature of the truncated Floquet--Magnus expansion.

\begin{figure}
 \centering
 \includegraphics[width=0.4\textwidth]{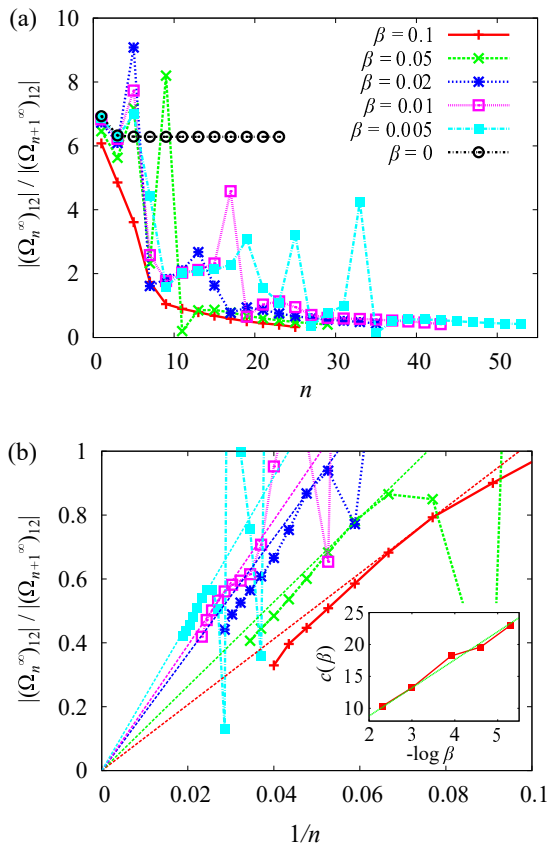}
 \caption{(a) Plot of $|(\Omega_n^{\infty})_{12} |/|(\Omega_{n+1}^{\infty})_{12} |$ as a function of $n$ for the driven anharmonic oscillator.
 The values of the anharmonic parameter $\beta$ are $0.1$, $0.05$, $0.02$, $0.01$, $0.005$, and $0$.
 The cutoff dimension is set to $D=64$.
 The dashed line of $\beta=0$ corresponds to the case of the harmonic oscillator.
 The values of other parameters are $\omega_0=g=1$.
(b) Same quantity as a function of $1/n$.
The thin lines represent the tangent lines that pass through the origin, whose slopes are denoted as $c(\beta)$.
The inset shows $c(\beta)$ as a function of $-\log \beta$.
The dashed line represent $c(\beta)=-\kappa \log \beta$ with $\kappa=4.4$.}
 \label{fig-Tc-AHO}
\end{figure}

We have numerically demonstrated that, for two types of the driven harmonic oscillators, $T_{\mathrm{c}}$ defined by Eq.~(\ref{Tc_ratio_infinity}) is identical to the radius of convergence obtained from the exact Floquet Hamiltonian.
These results justify Eq.~(\ref{Tc_ratio_infinity}) as the definition of the radius of convergence.
We now estimate $T_{\mathrm{c}}$ for the driven anharmonic oscillator defined by Eq.~(\ref{H_anharmonic}).
For simplicity, we employ the eigenstates of the harmonic oscillator as the orthogonal basis vectors to truncate the Hilbert space.
We can numerically confirm that $(\Omega_{n}^{\infty})_{ij}$ in Eq.~(\ref{Omega_infinity_limit}) exists for each index $(i,j)$.
The existence of this limit justifies the choice of the basis for the truncation.
Figure \ref{fig-Tc-AHO} (a) shows $|(\Omega_n^{\infty})_{12} |/|(\Omega_{n+1}^{\infty})_{12} |$ as a function of $n$ for different values of the anharmonic parameter $\beta$.
Each $|(\Omega_n^{\infty})_{12}|$ is calculated for a sufficiently large cutoff dimension $D$ so that it is well converged.
The dashed line of $\beta=0$ corresponds to the case of the harmonic oscillator and it has a constant value $T_{\mathrm{c}}=2\pi/\omega_0$.
In the numerical calculation of $(\Omega_{n}^{D})_{ij}$, the rounding error becomes severe as $n$ increases.
In Fig.~\ref{fig-Tc-AHO}, we have plotted up to the maximum $n$ below which the rounding error is negligible.
Figure \ref{fig-Tc-AHO} (b) shows $|(\Omega_n^{\infty})_{12} |/|(\Omega_{n+1}^{\infty})_{12} |$ as a function of $1/n$.
While for small $n$ the ratio $|(\Omega_n^{\infty})_{12} |/|(\Omega_{n+1}^{\infty})_{12} |$ shows an oscillating behavior, for sufficiently large $n$ it varies smoothly as a function of $n$.
We call the latter region of $n$ as an asymptotic regime.
Below, we show that for any $\beta>0$, $|(\Omega_n^{\infty})_{12} |/|(\Omega_{n+1}^{\infty})_{12} |$ vanishes in the limit $n \to \infty$.
In the asymptotic regime, let $c(\beta)$ be the slope of the tangent line that passes through the origin of Fig.~\ref{fig-Tc-AHO} (b).
These tangent lines are shown in the figure as thin dashed lines.
From Fig.~\ref{fig-Tc-AHO} (b), we have the following upper bound:
\begin{equation}
\frac{|(\Omega_{n}^{\infty})_{12}|}{|(\Omega_{n+1}^{\infty})_{12}|} < \frac{c(\beta)}{n},
\label{Omega_ratio_AHO_bound}
\end{equation}
in the asymptotic regime.
We assume that Eq.~(\ref{Omega_ratio_AHO_bound}) also holds for a larger $n$ that is not accessible in this numerical calculation.
The inset in Fig.~\ref{fig-Tc-AHO} (b) shows $c(\beta)$ as a function of $-\log \beta$.
This figure suggests that $c(\beta) = -\kappa \log \beta$ with $\kappa \simeq 4.4$, in other words, $c(\beta)$ is always finite for arbitrarily small $\beta>0$.
Thus, we can conclude that the radius of convergence $T_{\mathrm{c}}$ defined by Eq.~(\ref{Tc_ratio_infinity}) vanishes.

\section{Quantum resonance}
\label{sec:Quantum_resonance}

In Sec.~\ref{sec:Radius_of_convergence}, it has been shown that the Floquet--Magnus expansion for the driven anharmonic oscillator diverges for any $T>0$ and $\beta>0$.
We recall that the radius of convergence of the expansion is related to the resonance frequency.
However, in Sec.~\ref{sec:Driven_anharmonic_oscillator}, we have concluded that the energy of the driven anharmonic oscillator is bounded from above because the Floquet states are localized in energy space.
In this section, we attempt to resolve this paradox by arguing that there exist undetectable fine resonances arbitrarily close to $T=0$.

\begin{figure}
 \centering
 \includegraphics[width=0.4\textwidth]{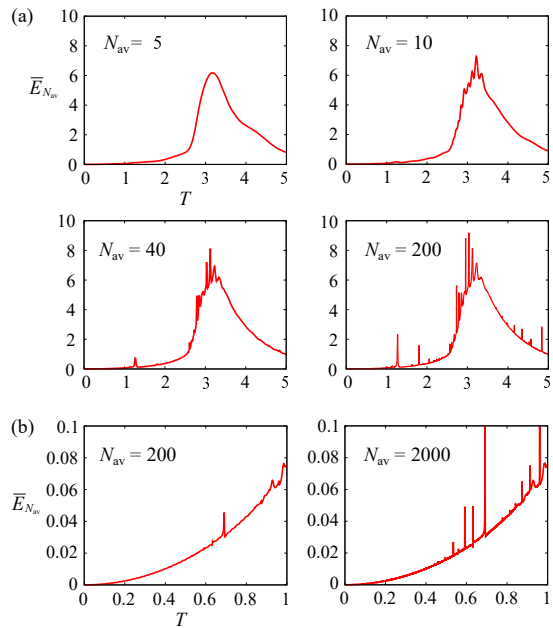}
 \caption{(a) Plot of $\bar{E}_{N_{\mathrm{av}}}$ as a function of $T$ for the driven anharmonic oscillator.
 The values of $N_{\mathrm{av}}$ are $5$, $10$, $40$, and $200$.
 The initial state is the ground state of the harmonic oscillator.
 The values of other parameters are $\omega_0=\beta=g=1$.
 The cutoff dimension is $D=64$.
 The expectation value for the initial state $\langle \psi_0| \hat{H}_0 |\psi_0 \rangle$ is subtracted from $\bar{E}_{N_{\mathrm{av}}}$ so that it vanishes at $T=0$.
 Many resonance peaks appear as $N_{\mathrm{av}}$ increases.
 (b) Enlarged plot around $T=0$.
 The values of $N_{\mathrm{av}}$ are $200$ and $2000$.}
 \label{fig-resonance-AHO}
\end{figure}

In Fig.~\ref{fig-E}, we have calculated the long-time average of the energy $\bar{E}$ as a function of the driving period $T$.
Here, we revisit this problem to investigate its fine structure that is not shown in Fig.~\ref{fig-E}.
To visualize the resonance structure of the driven anharmonic oscillator, we calculate the energy averaged over a finite time interval.
For a given initial state $|\psi_0\rangle$, the expectation value of the energy at $t=nT$ is written as
\begin{equation}
E(nT) = \langle \psi_0 | (\hat{U}^{\dag})^n \hat{H}_0 \hat{U}^n | \psi_0 \rangle.
\end{equation}
We define the averaged energy by
\begin{equation}
\bar{E}_{N_{\mathrm{av}}}(T) = \frac{1}{N_{\mathrm{av}}+1} \sum_{n=0}^{N_{\mathrm{av}}} E(nT).
\label{E_av_partial}
\end{equation}
While for a finite $N_{\mathrm{av}}$ the time-averaged energy $\bar{E}_{N_{\mathrm{av}}}(T)$ is an analytic function of $T$, the analyticity of its long-time limit $\bar{E}(T)=\lim_{N_{\mathrm{av}} \to \infty} \bar{E}_{N_{\mathrm{av}}}(T)$ is not guaranteed.
In the calculation of $\bar{E}_{N_{\mathrm{av}}}(T)$ by the exact diagonalization, the Hilbert space is truncated with respect to the eigenstates of the harmonic oscillator.
The cutoff dimension $D$ is chosen sufficiently large so that $\bar{E}_{N_{\mathrm{av}}}(T)$ is well converged.
Figure \ref{fig-resonance-AHO} shows $\bar{E}_{N_{\mathrm{av}}}(T)$ for different values of $N_{\mathrm{av}}$.
The initial state is the ground state of the harmonic oscillator.
Remarkably, as $N_{\mathrm{av}}$ increases, more and more resonance peaks appears. 
Figure \ref{fig-resonance-AHO} (b) is the enlarged version of (a).
One can see fine peaks in Fig.~\ref{fig-resonance-AHO} (b), which are invisible in (a) due to its poor resolution.
From these observations, we conjecture that infinitely many resonances appear arbitrarily close to $T=0$ as $N_{\mathrm{av}}$ further increases and the resolution of $T$ is improved.
This peculiar resonance structure provides a possible explanation for the divergence of the Floquet--Magnus expansion in the driven anharmonic oscillator.
It should be noted that, in real experiments, only a finite number of resonances are observable due to the limitation of the observation time and the frequency resolution.
The multiple resonances shown in Fig.~\ref{fig-resonance-AHO} do not exist in the classical driven anharmonic oscillator, and we thus call it the quantum resonance.
In the case of the driven harmonic oscillator ($\beta=0$), one can obtain the analytic form of $\bar{E}(T)$ that diverges at $T=2\pi/\omega_0$, and there does not exist such complicated resonance structure.

The existence of infinitely many resonances seems quite unusual because such behavior is never observed in classical systems.
Thus, we demonstrate that a simple quantum model also has the similar resonance structure.
One of the most frequently studied driven systems with energy localization is the kicked rotor model,
\begin{equation}
\hat{H}(t) = \frac{1}{2} \hat{p}^2 + K \cos \hat{x} \sum_{n=-\infty}^{\infty} \delta(t-nT),
\label{H_KR}
\end{equation}
where $K$ is the amplitude of the kick.
The time evolution operator for a single period is written as
\begin{equation}
\hat{U} = e^{-iT\hat{p}^2/2} e^{-iK\cos \hat{x}}.
\end{equation}
It is shown that, if $T/4\pi$ is an irrational number, the expectation value of the kinetic energy $\langle \hat{p}^2/2 \rangle$ remains finite, whereas if $T/4\pi$ is a rational number, it unboundedly increases as $\langle \hat{p}^2/2 \rangle \sim t^2$ (see Refs.~\cite{Izrailev-90} and \cite{Tian-10}).
This behavior can be explained as follows.
If one employs the eigenstates of the momentum $\hat{p}|n\rangle=n|n\rangle$ as the orthogonal basis vectors, the eigenstates of $\hat{U}$ (Floquet states) are identical to those of a one-dimensional tight-binding model with an inhomogeneous potential.
For an irrational $T/4\pi$, because this potential is nonperiodic, the Floquet states exhibit Anderson localization in the momentum space.
Thus, the expectation value of the energy remains finite.
In contrast, if $T/4\pi=p/q, (p,q=1,2,...)$, $\hat{U}$ has the following translational symmetry in the momentum space,
\begin{equation}
\langle m+q | \hat{U} | n+q \rangle = \langle m | \hat{U} | n \rangle.
\label{KR_trans_sym}
\end{equation}
According to Bloch's theorem, the modulus of the Floquet state is also a periodic function with period $q$.
Thus, each Floquet state is extended over the whole momentum space and the energy of the system increases indefinitely with time.
This resonance occurs even if the kick amplitude $K$ is arbitrarily small.
Note that such a resonant behavior is not observed in the classical kicked rotor, where there is a certain threshold $K_{\mathrm{c}}(T)$ for any $T$, such that the energy of the system remains finite if $K<K_{\mathrm{c}}(T)$.

\begin{figure}
 \centering
 \includegraphics[width=0.4\textwidth]{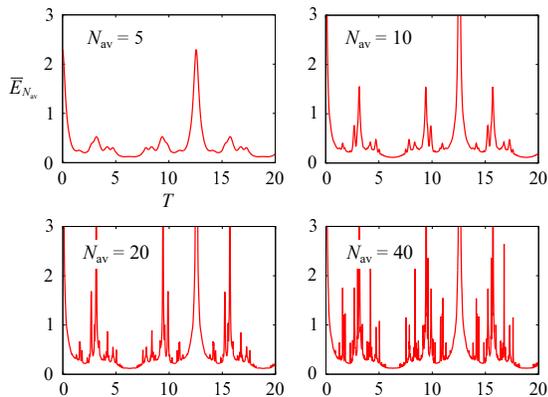}
 \caption{Plot of $\bar{E}_{N_{\mathrm{av}}}$ as a function of $T$ for the kicked rotor model.
 The values of $N_{\mathrm{av}}$ are $5$, $10$, $20$, and $40$.
 The initial state is the zero-momentum state.
 The amplitude of the kick is $K=1$.
 The cutoff dimension is $D=33$.
 Many resonance peaks appear as $N_{\mathrm{av}}$ increases.}
 \label{fig-resonance-KR}
\end{figure}

Let us visualize the resonance structure in the same manner as in Fig.~\ref{fig-resonance-AHO}.
We consider the average of the kinetic energy $E(nT) = \langle \psi_0 | (\hat{U}^{\dag})^n (\hat{p}^2/2) \hat{U}^n | \psi_0 \rangle$ over a finite time interval $N_{\mathrm{av}}$.
If $T/4\pi$ is an irrational number, the limit $\bar{E}(T) = \lim_{N_{\mathrm{av}} \to \infty} \bar{E}_{N_{\mathrm{av}}}(T)$ exists, and if $T/4\pi$ is a rational number, $\bar{E}_{N_{\mathrm{av}}}(T)$ diverges.
Thus, $\bar{E}$ is everywhere discontinuous as a function of $T$.
To calculate $\bar{E}_{N_{\mathrm{av}}}(T)$, the Hilbert space is truncated with respect to the momentum basis $\{ |n\rangle \}_{n=0,\pm1,...}$.
Figure \ref{fig-resonance-KR} shows $\bar{E}_{N_{\mathrm{av}}}$ as a function of $T$ for different values of $N_{\mathrm{av}}$.
The initial state is the zero-momentum state ($\hat{p}|0\rangle=0$).
As $N_{\mathrm{av}}$ increases, countless resonance peaks appear.
Each peak corresponds to a rational point $T/4\pi=p/q$ and the strongest resonances appear at $T/4\pi=0,1,2,...$.
As the denominator of $T/4\pi=p/q$ increases, the rate of the energy growth is suppressed exponentially \cite{Izrailev-90}.

The peculiarity of the quantum resonance in the kicked rotor is that infinitely many resonances are densely distributed over the whole $T$ region.
More precisely, for any $T$ and $\epsilon>0$, there exists $T' \in [T-\epsilon,T+\epsilon]$ such that $\lim_{N_{\mathrm{av}} \to \infty} \bar{E}_{N_{\mathrm{av}}}(T') = \infty$.
Thus, we conclude that, in the kicked rotor, the Floquet Hamiltonian as a continuous function of $T$ does not exist for all $T$.
Obviously, the radius of convergence of the Floquet--Magnus expansion vanishes.

In contrast to the kicked rotor, it is difficult to identify all resonances of the driven anharmonic oscillator because its Floquet operator $\hat{U}$ does not have any translational symmetry in energy space such as Eq.~(\ref{KR_trans_sym}).
At the present time, it is not clear whether the existence of infinitely many resonances is peculiar to the kicked rotor or a general feature of driven nonintegrable systems with energy localization.

\section{Concluding remarks}
\label{sec:Conclusions}

In this study, we have proposed a procedure to estimate the radius of convergence of the Floquet--Magnus expansion for periodically driven systems with an unbounded energy spectrum.
If the Hamiltonian is a bounded operator, the evaluation of the radius of convergence is straightforward.
In contrast, if the Hamiltonian is unbounded, there is ambiguity in the definition of the convergence of the expansion because the matrix norm is ill-defined.
In our procedure, we consider each matrix element of the Floquet--Magnus expansion corresponding to the Hamiltonian truncated up to a finite dimension.
By taking the limit of the cutoff dimension for each order of the expansion, we have defined the radius of convergence $T_{\mathrm{c}}$ by Eq.~(\ref{Tc_ratio_infinity}).
This definition can reproduce the analytical results for the exactly solvable models such as the driven harmonic oscillators.
By applying this procedure to the driven anharmonic oscillator given by Eq.~(\ref{H_anharmonic}), we have found that $T_{\mathrm{c}}$ vanishes even if the anharmonic coefficient $\beta$ is arbitrarily small.

We have also shown that the energy of the driven anharmonic oscillator remains finite and the level spacing of the quasi-energies follows the Poisson distribution.
In previous works \cite{D'Alessio-13,Lazarides-14,D'Alessio-14,Ponte-15-1,Moessner-17,Bukov-16}, the authors anticipated a direct connection between the divergence of the Floquet--Magnus expansion and the quantum ergodicity, which implies that each Floquet state is a linear combination of all available eigenstates of the unperturbed Hamiltonian and the Floquet operator exhibits properties of random matrices.
Our results contradict this consensus.
In Sec.~\ref{sec:Quantum_resonance}, we have proposed a possible explanation for the divergence of the expansion in the driven anharmonic oscillator.
From the careful investigation of the time-averaged energy, we conjectured that there exist infinitely many resonances arbitrarily close to $T=0$.
Since it is impossible to verify this conjecture numerically or experimentally owing to the limited resolution and observation time, some theoretical approaches are required to improve our understanding of this problem.

For periodically driven integrable systems such as the driven harmonic oscillators, the Floquet Hamiltonian can be calculated analytically and the Floquet--Magnus expansion is shown to converge for sufficiently high driving frequencies.
By generalizing the result of the driven anharmonic oscillator, we expect that arbitrary nonintegrable perturbations to integrable systems always lead to the divergence of the Floquet--Magnus expansion.
Currently, it is unclear whether this is also true in classical systems or if this is a peculiarity of quantum dynamics.

In this study, we have concentrate on one-body systems in a continuous space. 
It is also an interesting problem to consider the convergence of the Floquet--Magnus expansion for discrete many-body systems on a lattice, including spin chains and Hubbard models.
Since the dimension of the Hilbert space for these lattice systems becomes infinite in the thermodynamic limit, the truncation scheme developed in this study may be applicable to estimate the radius of convergence.
An important question is whether the Floquet--Magnus expansion converges for many-body localized systems.
It is known that disordered many-body systems cannot absorb energy from high-frequency driving if these systems exhibit many-body localization \cite{Ponte-15-2,Lazarides-15}.
Our result suggests the possibility that the Floquet--Magnus expansion diverges even if the system is in a strongly localized phase.

Another challenging problem is to understand the structure of quantum resonance in many-body systems with energy localization.
As an analytically tractable example of periodically driven many-body systems, the coupled kicked rotor model, where $N$ kicked rotors are fully connected by the kicking potential, was investigated in Ref.~\cite{Russomanno-18}.
It is analytically shown that, for $N\geq3$, there exists a threshold kick amplitude $K_{\mathrm{c}}$ above which the Floquet states are delocalized in energy space and the expectation value of the kinetic energy diverges in time.
For $K<K_{\mathrm{c}}$, the energy remains finite or diverges depending on whether $T/4\pi$ is an irrational or rational number, respectively.
The threshold $K_{\mathrm{c}}$ vanishes in the thermodynamic limit $N \to \infty$.
From the results of the coupled kicked rotor, one may be led to a conjecture that the existence of infinitely many quantum resonances discussed in Sec.~\ref{sec:Quantum_resonance} is a general feature of many-body systems with energy localization, such as disordered spin chains.

Although the Floquet--Magnus expansion diverges for the driven anharmonic oscillator, we expect that a finite truncation of the expansion provides approximate dynamics of the system.
More precisely, there is a timescale $\tau_*$ such that, at times $t<\tau_*$, the dynamics is approximatelly described by the effective Hamiltonian obtained by truncating the Floquet--Magnus expansion up to an optimal order $n_*$.
In fact, for nonintegrable many-body systems driven by a high-frequency field, the system first relaxes to a quasi-stationary state before reaching an infinite temperature state.
It has been shown theoretically that for discrete lattice systems with local interactions, the lifetime of the quasi-stationary state exponentially increases with the driving frequency, $\tau_* \sim e^{O(1/T)}$ \cite{Mori-16,Kuwahara-16,Abanin-17}.
This can be intuitively understood by noting that the absorption of the energy from the high-frequency driving requires the collective excitation of complex many-body states, which is exponentially suppressed in the frequency owing to the locality of the interaction.
The quasi-stationary state is described by the Gibbs state of the Floquet--Magnus effective Hamiltonian.
This behavior is known as ``Floquet prethermalization''.
The existence of a metastable state described by the effective Hamiltonian is also demonstrated for the periodically driven Friedrichs model with an unbounded energy spectrum, where the Floquet--Magnus expansion diverges for all $\omega$ \cite{Mori-15}.
In the context of our study, we ask how the timescale $\tau_*$ behaves as a function of the driving period $T$ and the anharmonicity $\beta$ in periodically driven few-body systems with energy localization.
This problem is closely related to the characterization of ``chaotic'' behavior in low-dimensional quantum systems.

\begin{acknowledgments}
The author thanks Shin-ichi Sasa for useful discussions.
The present study was supported by KAKENHI No.~JP17H01148.
\end{acknowledgments}

\appendix

\section{Floquet Hamiltonian for the driven harmonic oscillator}
\label{Appendix:driven_harmonic_oscillator}

In this Appendix, we derive Eq.~(\ref{HF_HO_analytic}) for the driven harmonic oscillator given by Eq.~(\ref{H_HO}).
By using the creation and annihilation operators,
\begin{equation}
\hat{a}^{\dag} = \sqrt{\frac{\omega_0}{2}} \biggl( \hat{x}-i\frac{\hat{p}}{\omega_0} \biggr), \:\:\: \hat{a} = \sqrt{\frac{\omega_0}{2}} \biggl( \hat{x}+i\frac{\hat{p}}{\omega_0} \biggr),
\end{equation}
the Hamiltonian Eq.~(\ref{H_HO}) is written as
\begin{equation}
\hat{H}(t) = \omega_0 \hat{a}^{\dag} \hat{a} + \lambda(t) \frac{g}{\sqrt{2\omega_0}} (\hat{a}^{\dag} + \hat{a}),
\end{equation}
where the zero point energy is omitted.
Recall that each term of the Floquet--Magnus expansion is represented in terms of the nested commutators of the Hamiltonian at different times.
In the case of the harmonic oscillator, only three operators $\hat{a}^{\dag} \hat{a}$, $\hat{a}^{\dag}$, and $\hat{a}$ appear in the expansion.
Thus, we write the Floquet Hamiltonian as
\begin{equation}
\hat{H}_F T = c_1 \hat{a}^{\dag} \hat{a} + c_2 \hat{a}^{\dag} + c_2^* \hat{a}.
\end{equation}
And then, the creation operator in the Heisenberg representation is calculated as
\begin{equation}
\hat{a}^{\dag}(T) = e^{i\hat{H}_F T} \hat{a}^{\dag} e^{-i\hat{H}_F T} = e^{ic_1} \hat{a}^{\dag} + \frac{c_2^*}{c_1} (e^{ic_1}-1),
\label{aT_Heisenberg_1}
\end{equation}
where we used Baker--Hausdorff formula,
\begin{equation}
e^{\hat{B}} \hat{A} e^{-\hat{B}} = \hat{A} + [\hat{B},\hat{A}] + \frac{1}{2!} [\hat{B},[\hat{B},\hat{A}]] + ...,
\end{equation}
and $[\hat{a},\hat{a}^{\dag}]=1$.

On the other hand, the time-evolution operator is also written as $\hat{U}=e^{-i\hat{H}_{-}T/2}e^{-i\hat{H}_{+}T/2}$, where $\hat{H}_{\pm} = \omega_0 \hat{a}^{\dag} \hat{a} \pm g/\sqrt{2\omega_0} (\hat{a}^{\dag} + \hat{a})$.
By using Baker--Hausdorff formula twice, $\hat{a}^{\dag}(T)$ is calculated as
\begin{eqnarray}
\hat{a}^{\dag}(T) &=& e^{i\hat{H}_{+}T/2} e^{i\hat{H}_{-}T/2} \hat{a}^{\dag} e^{-i\hat{H}_{-}T/2} e^{-i\hat{H}_{+}T/2} \nonumber \\
&=& e^{i\omega_0 T/2} e^{i\hat{H}_{+}T/2} \hat{a}^{\dag} e^{-i\hat{H}_{+}T/2} + \frac{g}{\sqrt{2\omega_0}} \frac{1-e^{i\omega_0 T/2}}{\omega_0} \nonumber \\
&=& e^{i\omega_0 T} \hat{a}^{\dag} + \frac{g}{\sqrt{2\omega_0}} \frac{(1-e^{i\omega_0 T/2})^2}{\omega_0}.
\label{aT_Heisenberg_2}
\end{eqnarray}
Comparing Eqs.~(\ref{aT_Heisenberg_1}) and (\ref{aT_Heisenberg_2}), we find
\begin{eqnarray}
c_1 = \omega_0 T, \:\:\: c_2 = -\frac{i g}{\sqrt{2\omega_0}} T \tan \frac{\omega_0 T}{4},
\end{eqnarray}
thus we have Eq.~(\ref{HF_HO_analytic}).

\section{Numerical method to calculate the Floquet--Magnus expansion}
\label{Appendix:Floquet_Magnus_expansion}

In this Appendix, we explain the numerical scheme to calculate higher-order terms of the Floquet--Magnus expansion.
If we employ the protocol given by Eq.~(\ref{lambda_step}), the time-evolution operator is written as
\begin{equation}
U = e^{tX} e^{tY},
\end{equation}
where $U$, $X$, and $Y$ are $D$ dimensional matrices.
$U$ is expanded with respect to $t$ as
\begin{eqnarray}
U = I + \sum_{n=1}^{\infty} P_n, \:\:\: P_n = t^n \sum_{k=0}^{n} \frac{X^{n-k} Y^k}{(n-k)! k!}.
\end{eqnarray}
If we define a matrix $R$ by $U=e^{R}$, it is expanded as
\begin{equation}
R=\sum_{n=1}^{\infty} R_n,
\label{R_expansion}
\end{equation} 
where $R_n=O(t^n)$ is given by the so-called Baker--Campbell--Hausdorff formula.
For example, the first four terms read
\begin{eqnarray}
R_1 &=& t(X+Y), \nonumber \\
R_2 &=& \frac{1}{2} t^2 [X,Y], \nonumber \\
R_3 &=& \frac{1}{12} t^3 ([X,[X,Y]]+[Y,[Y,X]]), \nonumber \\
R_4 &=& - \frac{1}{24} t^4 [Y,[X,[X,Y]]].
\end{eqnarray}
However, this formula is not convenient for numerical calculations because each $R_n$ becomes exponentially complicated as the order of the expansion increases.
Thus, we employ the following recursive scheme \cite{Klarsfeld-89}.
In terms of $R_n$, each $P_n$ can be written as
\begin{eqnarray}
P_1 &=& R_1, \nonumber \\
P_2 &=& R_2 + \frac{1}{2} (R_1)^2, \nonumber \\
P_3 &=& R_3 + \frac{1}{2}(R_1 R_2 + R_2 R_1) + \frac{1}{3!} (R_1)^3,
\end{eqnarray}
in general form,
\begin{eqnarray}
R_n &=& P_n - \sum_{m=2}^n \frac{1}{m!} Q_n^{(m)}, \nonumber \\
Q_n^{(m)} &=& \sum_{i_1+...+i_m=n} R_{i_1} ... R_{i_m},
\end{eqnarray}
where $Q_n^{(n)}=(R_1)^n$ and $Q_n^{(1)}=R_n$.
Here, $Q_n^{(m)}$ can be calculated recursively as follows,
\begin{equation}
Q_n^{(m)} = \sum_{l=1}^{n-m+1} R_l Q_{n-l}^{(m-1)}, \:\:\: (2 \leq m \leq n).
\end{equation}


\end{document}